\documentstyle[11pt,pasp,twoside,psfig]{article}
\def\ls{{_<\atop^{\sim}}}
\def\gs{{_>\atop^{\sim}}}

\def\edcomment#1{\iffalse\marginpar{\raggedright\sl#1\/}\else\relax\fi}
\reversemarginpar

\begin{document}
\title{Cosmology studies with GRB afterglows}
 \author{Fabrizio Fiore}
\affil{Osservatorio Astronomico di Roma, Monteporzio 
I00040 Italy}

\begin{abstract}
GRB afterglows close to their peak intensity are among the brightest
sources in the sky. 
We discuss how GRB optical-to-X-ray afterglows can
be used as probes of the metal enrichment of the interstellar matter
in the GRB host galaxies and of the heating history of the
intergalactic matter in filaments and outskirts of clusters of
galaxies along the line of sight.  The key points of the proposed
observation strategy are: 1) prompt GRB observations (from minutes to
a few hours); 2) coordinated high resolution O-UV and X-ray observations
(R=$1500-6000$ in the 0.1-1 keV band and R$\gs10,000$ in the OUV).
\end{abstract}

\section{Introduction}

Soon after the $\gamma$-ray flash, optical and X-ray afterglows of
GRBs are among the brightest sources in the sky at cosmological
redshifts.  Eighteen GRB redshifts have been measured to date, in the
range 0.0085--4.5. Follow-up observations of the three brightest GRB
localized by BeppoSAX (GRB990123, GRB990510 and GRB010222) show that
tens of minutes after the GRB the optical afterglow can be as bright
as 14-16 mag.; a few hours later it can still be as bright as 17-19
mag.  The X-rays afterglow can be as bright as the Crab Nebula a few
minutes after the GRB, while 5-8 hours later it can be as bright as a
bright AGN, i.e. $\approx$mCrabs.  (see Fiore et al. 2000 for an
estimate of the GRB afterglows 0.5-2 keV logN-logF, when fluxes are
integrated from minutes up to hours after the GRB event).  This opens
up a new perspective for gathering spectra of unprecedent quality of
sources at cosmological redshift, provided that the afterglow can be
observed in such short timescales.  Afterglows can then be used as
probes of the high z Universe through the detection of absorption line
systems along the line of sight, both intervening and associated with
the host galaxies.  Absorption lines are more difficult to study than
emission lines, but carry with them unbeatable information because
they probe matter along a single beam, i.e. along the line of sight to
the background beacon.  This greatly reduces the complications related
to the matter geometry and dynamics, which strongly affects emission
line studies. At least two fundamental issues can be addressed through
absorption line studies: (1) The enrichment history of the Universe,
through the study of the inter-stellar matter (ISM) of the GRB host
galaxies; (2) the heating history of the Universe, through the study
of the so called ``X-ray forest'' (high ionization resonant lines of
O, Ne, C, etc.) associated with the warm-hot phase of the
inter-galactic matter (IGM) in filaments and the outskirts of clusters
of galaxies along the line of sight (Fiore et al. 2001).

\section{Enrichment in high redshift galaxies}

The study of z$\gs1$ galaxies available so far relies on the
Lyman-break galaxies at z$=3-4$ (see e.g. Steidel et al.  1999) 
and on galaxies which happens to be along the line
of sight to bright background quasars (see e.g. Churchill et
al. 2000). Some of these systems are associated with Damped Ly$\alpha$
systems (DLA) (see e.g.  Pettini et al., 1997, 1999).  However,
Lyman-break galaxies are characterized by pronounced star-formation
and their inferred chemical abundances may be related to these regions
rather than being representative of typical high z galaxies.  Metal
line systems in galaxies along the line of sight to quasars probe
mainly the ISM of outer haloes, rather than that in the bulge or in
the disc.  Finally, it is not clear if galaxies associated with DLAs
are truly representative of the whole high z galaxy population. GRB
afterglows can provide new, independent  and less biased
tools to study the ISM of high z galaxies.

\begin{figure}[t!]
\centerline{
\hbox{
\psfig{file=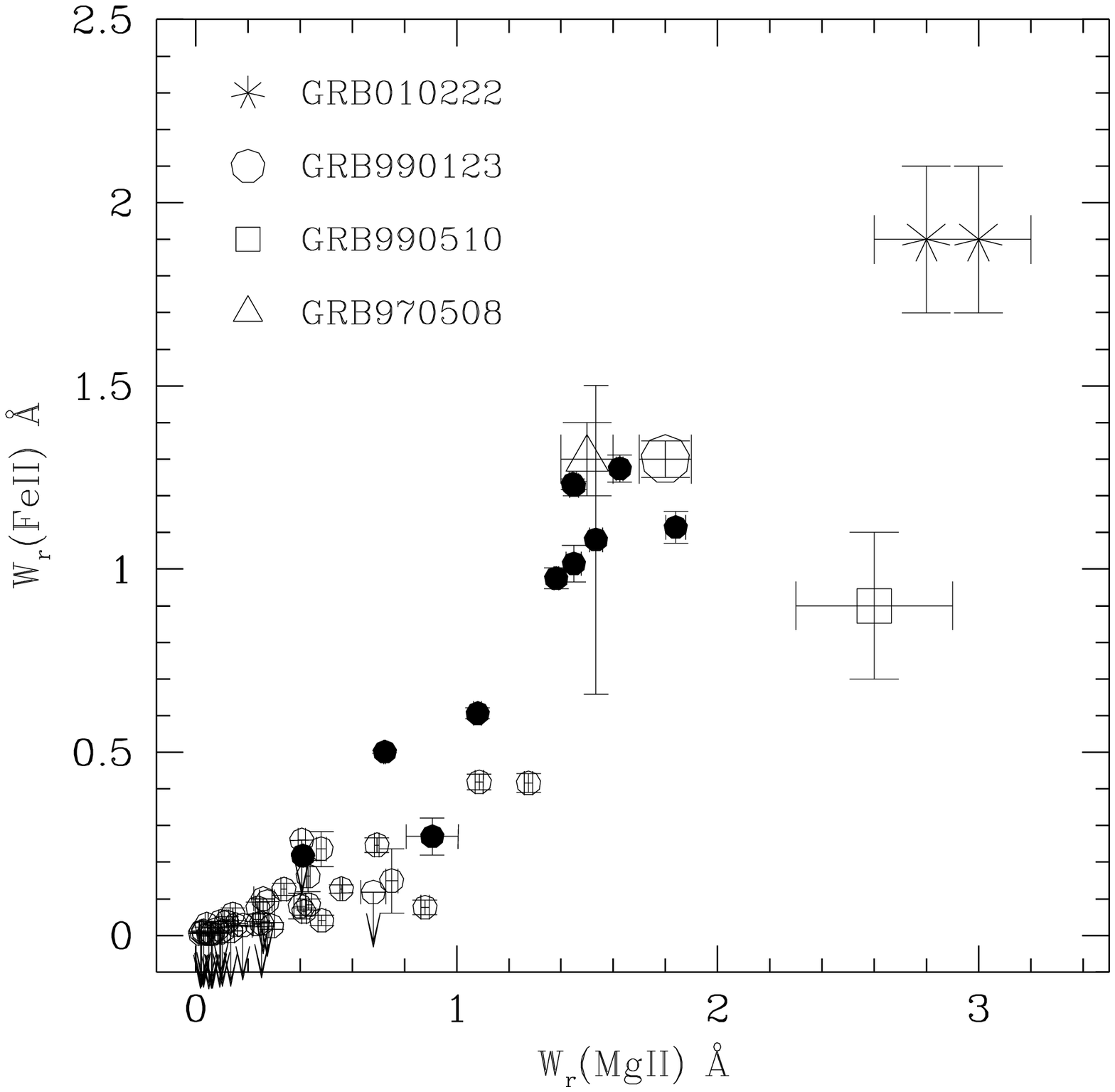,width=6.5cm,height=6.5cm,angle=0}
\psfig{file=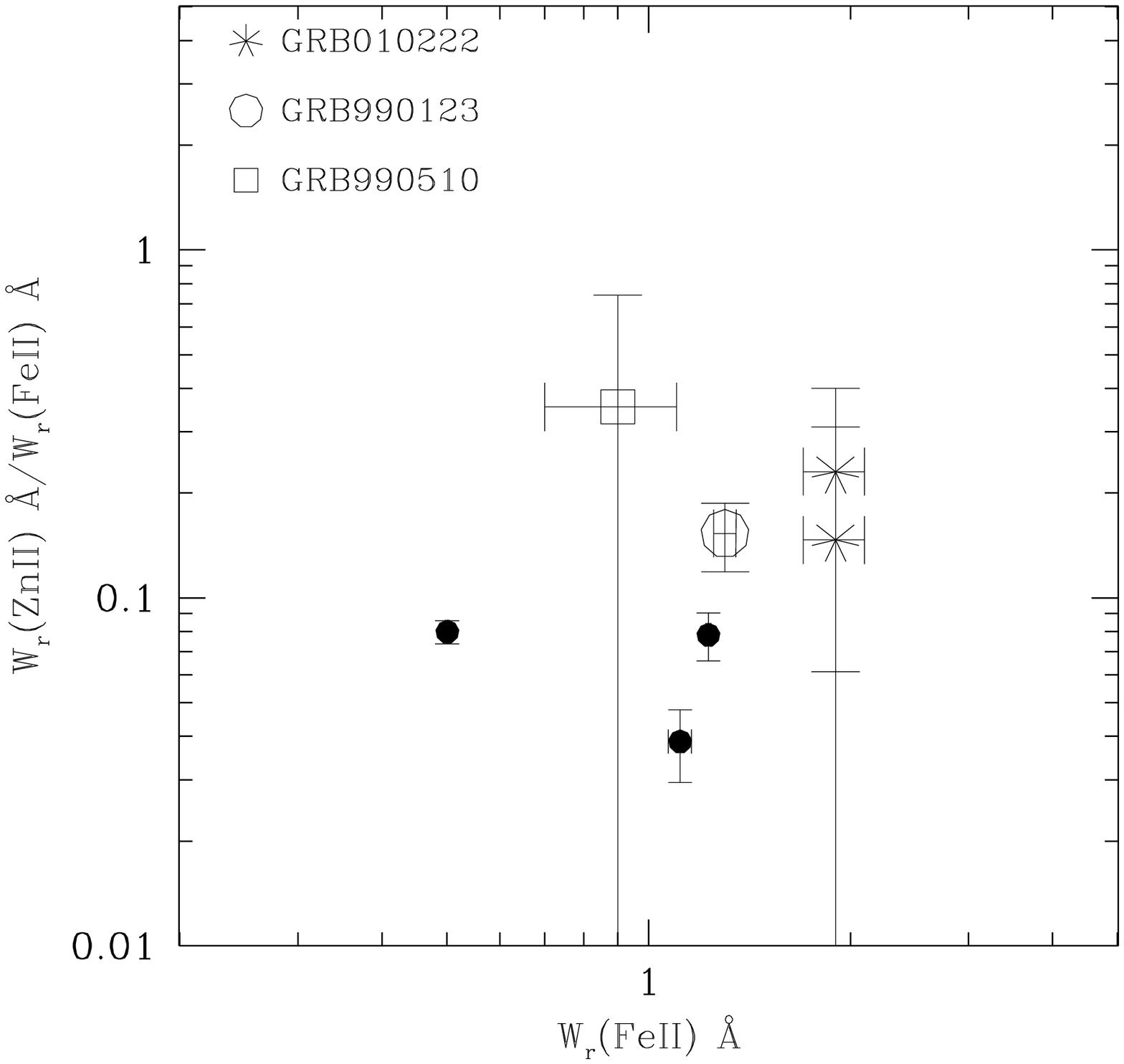,width=6.5cm,height=6.5cm,angle=0}
}
}
\caption{a) The EW of the MgII2796\AA~ and Fe2600\AA~ lines detected in the
spectra of four GRB afterglows: large symbols (from Masetti et
al. 2001, Jha et al. 2001, Vreeswijk et al. 2001, Kulkarni et al. 1999,
Metzger et al. 1997).  The MgII and FeII
line EW from the samples of Churchill et al. (2000), Lu et al. (1996) and
Pettini et al. (1999); small symbols. Metal lines associated with DLAs: 
filled circles. b) the ZnII2062 to FeII2600\AA~ EW ratios for
three DLAs studied by Lu et al. 1996 and Pettini et al. 1999:
filled circles. ZnII/FeII EW ratios of three GRB host galaxies: large
symbols. The ZnII2062 line is assumed to be $\approx40\%$ of the 
observed ZnII2062+CrII2062.6 line blend EW.}
\end{figure}

Figure 1a compares the equivalent width (EW) of the MgII2796\AA~ and
FeII2600\AA~ lines detected in the spectra of four GRB afterglows, and
probably associated with the host galaxies, with those in the samples
of Churchill et al. (2000), Lu et al. (1996) and Pettini et
al. (1999).  The EW of both lines is consistent or slightly higher (a
factor of 2 at most) than those of the strongest metal lines
associated with DLAs.  Figure 1b compares the ZnII2062\AA~ to
FeII2600\AA~ EW line ratios of three DLAs to those of three GRB host
galaxies.  Since the low resolution GRB afterglow observations cannot
resolve the ZnII2062.0 line from the nearby CrII2061.6 line we have
evaluated the relative contribution of the two lines to the observed
EW. The oscillator strength of the ZnII line is 3.3 times that of the
CrII line. If the Zn/Cr ratio in the GRB host galaxies is
$\approx0.2$, the mean Zn/Cr ratio in DLAs (Pettini et al. 1997), we
expect that the ZnII EW is $\approx40\%$ of the measured ZnII+CrII
blend. If this is the case GRB ZnII/FeII line rations could still be a
factor of 2-3 higher than DLA ones. This may imply a different dust
depletion pattern (Fe tends to be heavily locked in dust grain, while
Zn is almost unaffected by dust).  It is clear that only high
resolution spectra (R$\gs10000$) can provide unambiguous information
on the interstellar gas metallicity. 
Fe locked in dust grains may leave a signature in the
X-ray spectrum of GRB afterglows. Neutral Fe L edges have recently
been detected in a high resolution spectrum (R$\sim1000$) of an AGN
(Lee et al. 2001). The simultaneous absence of strong 
low ionization O and Ne edges can
be explained if most neutral Fe is locked in dust grains.  This
suggests that high resolution X-ray spectra of GRB afterglows can be
used to search for the signatures of Fe locked in dust in the host
galaxy ISM. X-ray spectroscopy may then represent a powerful tool,
complementing high resolution O-UV observations, in depicting the
chemical evolutionary history of galaxy ISM. This can then be compared
to the cosmic history of the metal enrichment derived through
different techniques (e.g. DLA studies) and to theoretical models (see
e.g. Cen \& Ostriker 1999).  The ISM of
GRB host galaxies may have already been detected in X-rays. Absorption
at a level of a few$\times10^{21}$ cm$^{-2}$ (rest frame) has been
discovered in the X-ray spectra of GRB010222 (in 't Zand et al.  2001,
Stratta et al. 2001 in preparation). Unfortunately the
low resolution (R=10) BeppoSAX spectra cannot detect individual
absorption edges and therefore the detailed study of GRB host galaxies
ISM must await for XMM-Newton, Chandra (see e.g. Piro et al. 2000),
or future missions R$\gs1000$ spectra.

Strong lines from $\alpha-$process elements like S and Si are
considered a signature of TypeII supernova enrichment. A SiII1526.7
line has been measured for the first time in the spectrum of a GRB
afterglow only recently (GRB010222, Masetti et al. 2001) with
EW=$1.2\pm$0.3\AA~.  This is in the ballpark of the EW of the SiII
lines associated with DLAs.

\section{The warm intergalactic matter}

According to current cosmological scenarios (see e.g. Dav\`e et
al. 2001) the warm ($10^5$ K$<$T$<10^7$ K) IGM may comprise $30-40\%$
of the baryons in the Universe at z$<1$.  Observations of the warm IGM
predicted to be found far from the high density virialized regions
have yielded so far only limited information. Gas with T$\approx10^5$
K has been revealed through OVI absorption at 1032, 1038\AA~ at
z=0.1--0.3 (e.g. Tripp et al.  2000).  In collisional equilibrium OVI
is the dominant O ion only for a very narrow temperature range
(T=$1-3\times10^{5}$ K). At higher temperatures the dominant ions are
OVII and OVIII, whose K$\alpha$ and K$\beta$ lines are in the range
0.5-0.7 keV.  X-ray spectroscopy of these lines (and of the high
ionization C,Ne, S and Si lines) is therefore a powerful tool to
probe such a warm, low density gas.  The expected EW for warm
intergalactic clouds of H column density between $10^{19}$ and
$10^{20}$ cm$^{-2}$ and metal abundance 0.2-0.3 solar are $\ls1$ eV,
calling for high resolution (R=$>1000$) spectroscopy.  This is
feasible today only with dispersive spectrometers (Elvis, these
proceedings).  One interesting property of gratings is that their
resolution is about constant in wavelengths and therefore it improves
as $E^{-1}$ going toward low energies. The minimum detectable EW
scales as the square root of the resolving power $\Delta E$.  Since
the rest frame line EW scales as (1+z)$\times EW_{obs}$, using
gratings the minimum detectable rest frame EW is about constant with
z. This means that similar O column densities can be probed up to the
z when the O features go out of the observing band (0.2 keV for
CCDs, corresponding to z$\approx2$ for OVIII).

Thermal broadening of O lines is $\sim50$ km s$^{-1}$ at
T=$4\times10^6$ K. A resolution of R$\gs6000$ is needed to resolve
these lines. Reasonable warm intergalactic gas turbulence may be of
100-200 km s$^{-1}$. In these cases a resolution of 1500-3000 may
still be enough to resolve the lines and measure the Doppler term
$b$. If the temperature of the gas can be constrained through OVI,
OVII and OVIII line ratios the measure of $b$ can provide information
on the heating history of the gas. For example, if the gas was shock
heated during the collapse of density perturbation one would expect
that the gas temperature is proportional to the square of the gas
sound speed, which in turn should be proportional to the gas
turbulence. By measuring $b$ and T it would be possible to perform a
self-consistency check of this idea and to provide tests and
constraints to hydrodynamic models for the collapse of density
perturbation and their evolution.

%

We thank A. Antonelli, E. Branchini, M. Elvis, A. Fontana,
E. Giallongo, N. Menci, F. Nicastro, C. Norman, S. Savaglio,
L. Stella, G. Stratta and M. Vietri, for useful discussions.

\end{document}